\documentclass[11pt,twoside]{article}
\usepackage{graphicx,epsfig,natbib,epstopdf}
\usepackage{CS18}
\begin{document}
\title{Investigating Brown Dwarf Variability at 3.4 \& 4.6$\mu$m with AllWISE Multi-Epoch Photometry}
\author{Gregory N.\ Mace\altaffilmark{1,2}  
}

\affil{$^1$Department of Physics and Astronomy, UCLA, 430 Portola Plaza, Box 951547, Los Angeles, CA, 90095-1547, USA}
\affil{$^2$Department of Astronomy, The University of Texas at Austin, 2515 Speedway, Stop C1400, Austin, TX 78712, USA; gmace@astro.as.utexas.edu}

\begin{abstract}
Multi-epoch photometry from AllWISE provides the opportunity to investigate variability at 3.4 and 4.6$\mu$m for most known brown dwarfs. 
WISE observed the same patch of sky repeatedly and within a day's time, roughly 12 observations were obtained on a given patch of sky; 
then, another 12 were obtained roughly six months later when that patch of sky was again in view. 
For most of the sky, AllWISE contains two separate epochs of about a dozen observations each, although $\sim$30\% of the sky has three such epochs available in AllWISE.  
With the AllWISE multi-epoch photometry of $\sim$1500 known M, L, T, and Y dwarfs, I computed the Stetson J Index and quantified variability as a function of spectral type. 
I found that the average single-exposure photometric uncertainty in AllWISE ($\sim$0.2 magnitudes) is too large to robustly identify flux variability smaller than $\sim$20\%.
However, multi-epoch photometry from AllWISE remains a useful resource in cases where flux variability is known to be present with large amplitudes, or for bright nearby objects with lower photometric uncertainties.
\end{abstract}

\section{Summary}
The study of brown dwarf \index{Brown Dwarf} flux variability \index{variability} has progressed rapidly. \index{WISE}
Early observations of variability proved its existence \citep{bailerjones2001, gelino2002, clarke2008}, but more recent studies have measured periodicity, wavelength/pressure dependencies, and long term evolution \citep{artigau2009,radigan2012,radigan2014,buenzli2012,buenzli2014,buenzli2014b,gillon2013,gizis2013,biller2013,apai2013,crossfield2014,burgasser2014, wilson2014}.
Although clouds are the most likely culprits of short-term variability \citep[][and references therein]{morley2014} on the order of a rotation period \citep[2-12hours;][]{reiners2008,artigau2009}, long-term variability caused by atmospheric circulation \citep{showman2013, zhang2014} or thermal perturbations \citep{robinson2014} is also possible. 
Because the atmospheres of brown dwarfs are accessible proxies for exoplanets, there is an increasing need for high-precision studies that combine multi-wavelength photometry and spectroscopy (see others proceedings from this conference).
Future studies will need to focus on a few bright and interesting objects to clarify the details, but the broader population must also be considered. 

The primary motivation for searching AllWISE photometry for variability was to inspect brown dwarf variability at 3.4 and 4.6$\mu$m with a statistical approach. 
To do this, I first compiled a census of $\sim$1850 known M, L, T, and Y dwarfs from DwarfArchives.org and the literature through 2014 February. 
I then employed the AllWISE Multiepoch Catalog, which is a compilation of $\sim$13 months of photometric observations from the Wide-field Infrared Survey Explorer \citep[WISE; ][]{wright2010}, to compute the Stetson J Index for each object \citep{stetson1996}. 
In these proceedings I highlight my analysis and attempt to guide other researchers in their use of AllWISE multi-epoch photometry for thermal infrared variability studies.

I find that the average single-exposure photometric uncertainties in AllWISE are $\sim$0.2 mag, which is larger than most of the variability measured in the literature.
As a result, the robust identification of variability in all brown dwarfs is not possible with the AllWISE Multiepoch Catalog. 
With the addition of phase information, provided by other near-infrared observations, one might tease out reliable signatures of variability at 3.4 \& 4.6$\mu$m. 
However, the AllWISE sampling of most sources is sparse enough to prevent robust periodogram analysis.
A few objects with measurable variability, and a couple objects of interest from the literature that were excluded from my analysis, would be good follow-up targets for future variability studies.
I recommend that other variability studies of brown dwarfs inspect the multi-epoch photometry in AllWISE \citep[and the ongoing NEOWISE-R mission; ][]{wright2014} to at least provide limits on the thermal infrared variability.
Additional investigation of AllWISE multi-epoch photometry with the Welch-Stetson Index may identify correlated variability between the W1 and W2 bands, and would be a useful followup analysis to what I present here.

\section{AllWISE Multi-Epoch Photometry for Brown Dwarfs}

The initial catalogs produced from the WISE mission were separated into the cryogenic and the post-cryogenic phases.
In both phases of the mission, the 3.4 and 4.6$\mu$m (W1 and W2) bands remained fully functional, while the 12 and 22 $\mu$m (W3 and W4) bands could not be used once cryogen was depleted.
The AllWISE processing of the WISE photometry combines the single-exposure images from the entire mission, between 2010 January and 2011 February, to improve co-added photometric measurements and provide uniform multi-epoch photometric and astrometric measurements \citep{cutri2013}.
The astrometric measurements in AllWISE are combined to provide the apparent motion (combined parallax and proper motion) of each detection \citep{wright2014,kirkpatrick2014}. 
There is also a four digit variability flag in the WISE and AllWISE catalogs that has been derived by comparing the dispersion in each objects multi-epoch photometry to the dispersion of the background sources \citep{hoffman2012}. The variability flag does not use the traditional Stetson L, J, and K indices \citep{welch1993, stetson1996}.

\begin{figure}[!htb]
\centering
\includegraphics[scale=0.52,angle=0]{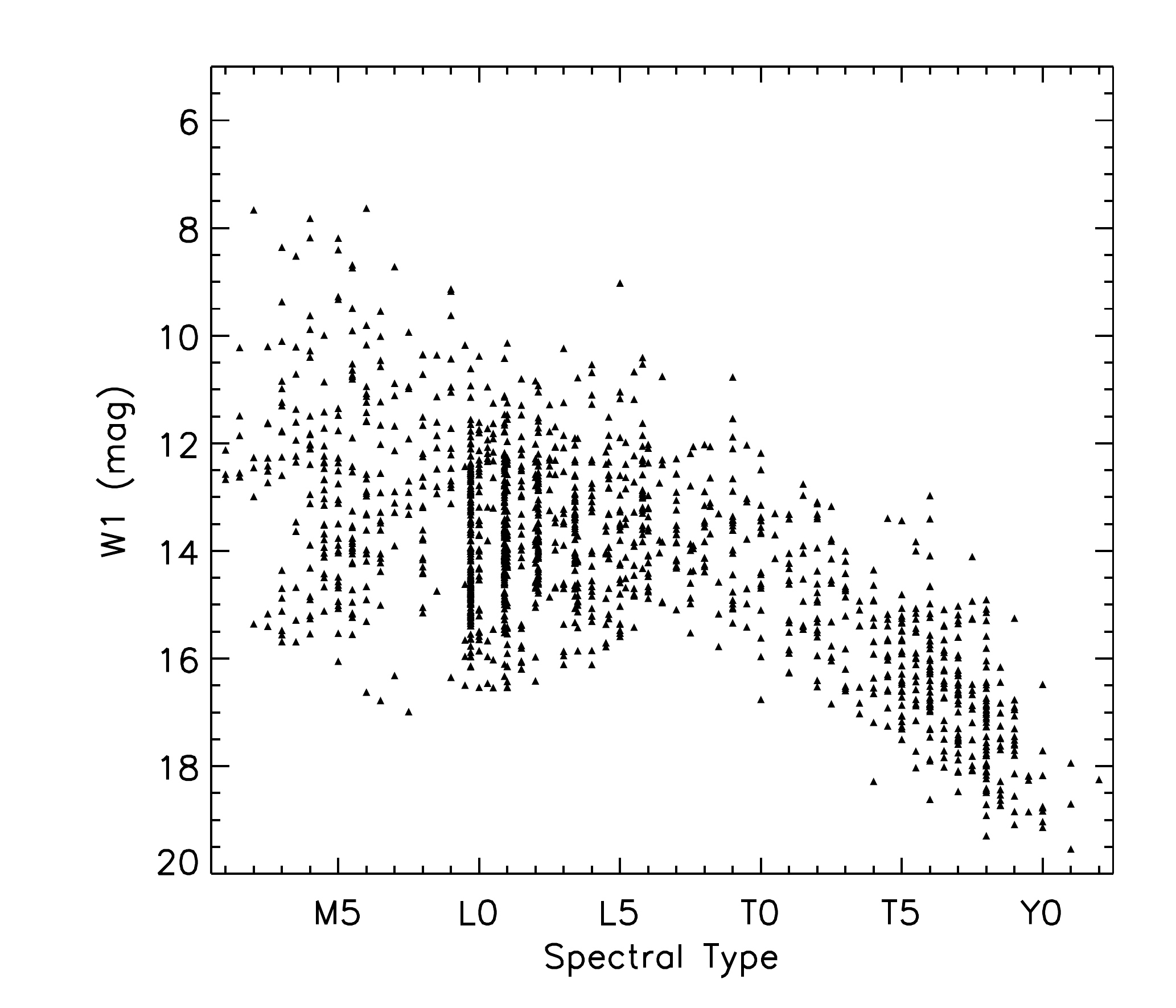}
\includegraphics[scale=0.52,angle=0]{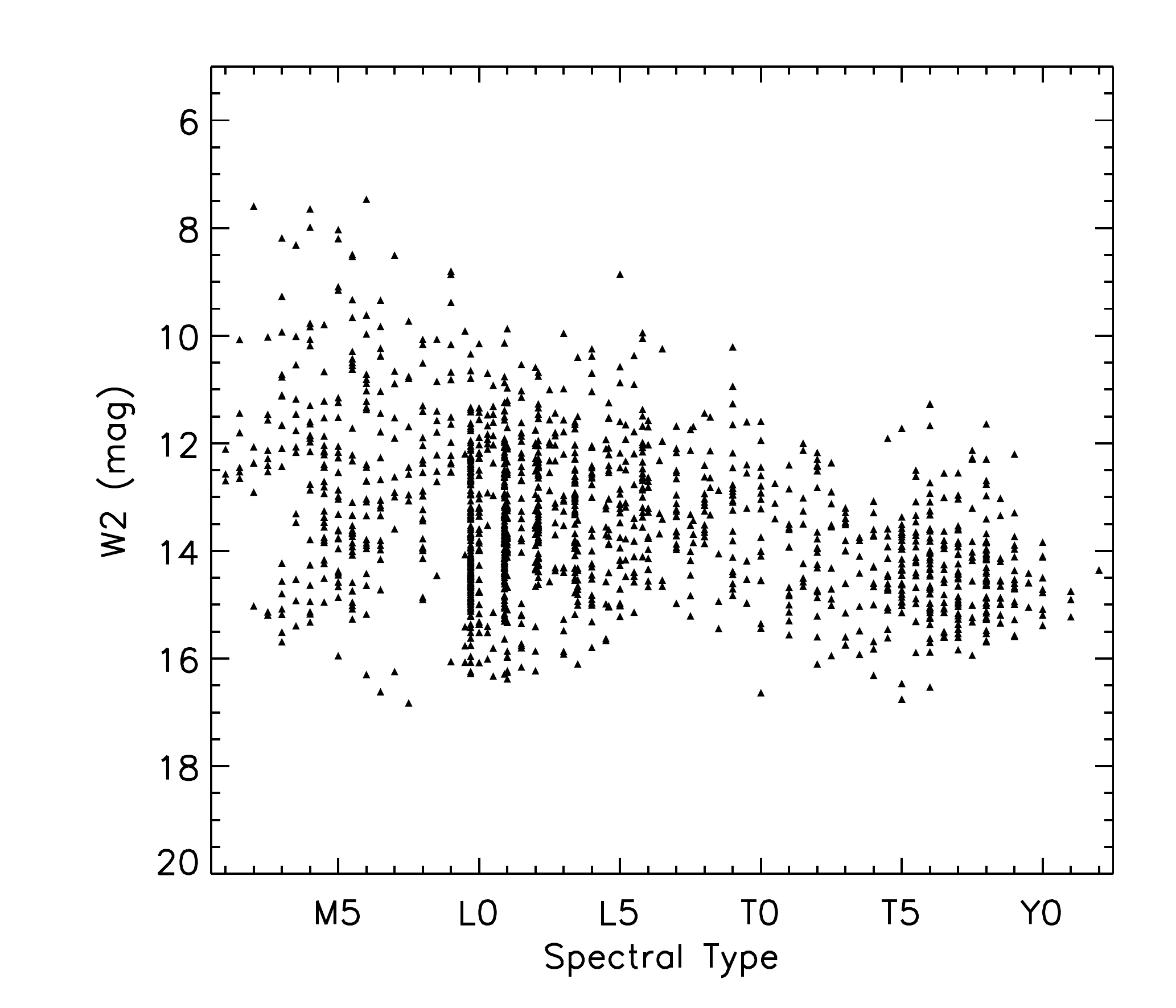}
\caption{AllWISE co-added W1 and W2 magnitudes as a function of spectral type. Uncertainties are not displayed because they are typically smaller than the points shown.
\label{fig1}}
\normalsize
\end{figure}
\clearpage

The M, L, T, and Y dwarf sample for this study started with the 2012 November version of DwarfArchives.org and was then updated for discoveries through 2014 February.
Figure~\ref{fig1} shows the co-added W1 and W2 magnitudes as a function of spectral type for the entire $\sim$1850 objects in the sample.
M dwarf spectral types are those taken from DwarfArchives, and are based on optical spectroscopy.
The L, T, and Y dwarf spectral types are all near-infrared types. L dwarf optical spectral types were converted to near-infrared types by L$_{opt}$ = 0.82*(L$_{IR}$) + 0.25, which was derived from the $\sim$90 L dwarfs with both optical and near-infrared spectral types in DwarfArchives.
The average photometric uncertainty for co-added AllWISE W1 and W2 photometry is very low ($\sim$0.03 mag, smaller than the points in Figure~\ref{fig1}) and only about 3 times higher for the faintest brown dwarf detections. In the W3 and W4 filters the co-added photometric uncertainties are much higher, $\sim$0.24 and $\sim$0.4 mag, respectively.

Not every object in the initial sample of $\sim$1850 objects is suited for studying variability.
The two primary issues are saturation of the early-type brown dwarfs and PSF blending with background sources.
To avoid these issues I imposed the following selection criteria:\\

1. The object must be in both the WISE All-Sky and AllWISE Source Catalogs. 
Objects that fail to make it into both catalogs are generally faint and close to other sources.
In the six-month time interval between WISE epochs, sources have moved enough to become (un)blended with a neighbor. 
An example of this is WISEPA J154151.66$-$225025.2 \citep[Y0.5; ][]{cushing2011}, which becomes blended at later epochs and is not detected by AllWISE.\\

2. Photometric measurements from the AllWISE Multiepoch Catalog are required to have detections with SNR $>$ 2.\\

3. The extended source flag ($ext\_flg$) must equal 0 in both the All-Sky and AllWISE Catalogs. 
This ensures that the source shape is consistent with a point source, ruling out blending.\\

4. The source was fit and measured using a single PSF component ($nb=1$) in both the All-Sky and AllWISE catalogs.\\

5. The saturation flags (w1sat, w2sat) are 0, which means that no pixels are saturated in the single-exposure photometry.
The single-exposure saturation limit is W1$<$8 and W2$<$7 magnitudes \citep{cutri2013}.\\

Of the original $\sim$1850 objects, 1510 make it through my selection process. 
Approximately 50 of the culled objects have no WISE detection and are late-type T dwarfs from UKIDSS.
I inspected the W3 and W4 photometry in the same manner as the W1 and W2 bandpasses, but they have large uncertainties and minimal time coverage. 
I exclude W3 and W4 from further discussion because none of the brown dwarfs are identified as variable in these passbands.

\section{Measuring Variability with the Stetson J Index}

The Stetson J Index was developed as a tool to identify photometric variables by weighting the difference in two photometric measurements by the time interval between the observations \citep{stetson1996}.
Short cadence observations with large amplitude changes give larger index values than smaller amplitude variations on the same timescale.
As applied by \citet{zhang2003}, the Stetson J Index is computed from the magnitude residual of two photometric measurements, 

\begin{figure}[!ht]
\centering
\includegraphics*[clip,trim=0mm 0mm 0mm 0mm,scale=0.2,angle=0]{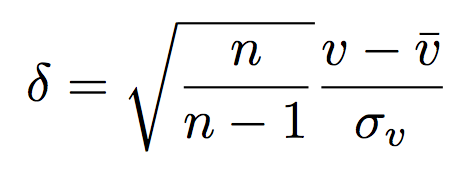}
\normalsize
\end{figure}

which are multiplied by each other,

\begin{figure}[!ht]
\centering
\includegraphics*[clip,trim=0mm 0mm 0mm 0mm,scale=0.2,angle=0]{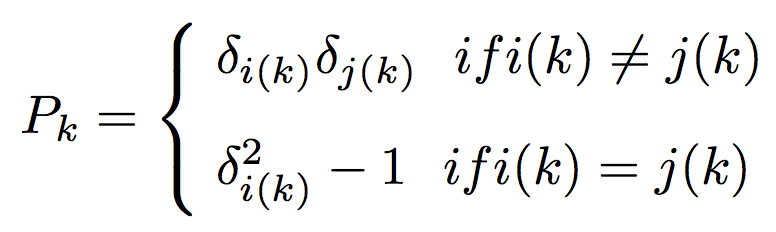}
\normalsize
\end{figure}

and then weighted by the time between the observations,\\

\begin{figure}[!hbt]
\centering
\includegraphics*[clip,trim=0mm 0mm 0mm 5mm,scale=0.3,angle=0]{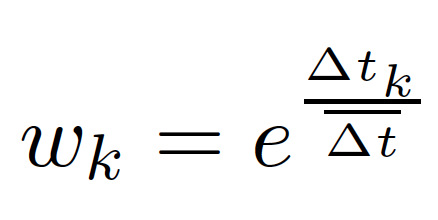}
\normalsize
\end{figure}

The index is computed as the normalized sum of the weighted pairs.

\begin{figure}[!ht]
\centering
\includegraphics*[clip,trim=0mm 0mm 0mm 0mm,scale=0.2,angle=0]{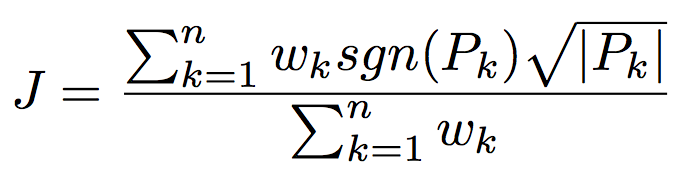}
\normalsize
\end{figure}

\begin{figure}[!ht]
\centering
\includegraphics[scale=0.52,angle=0]{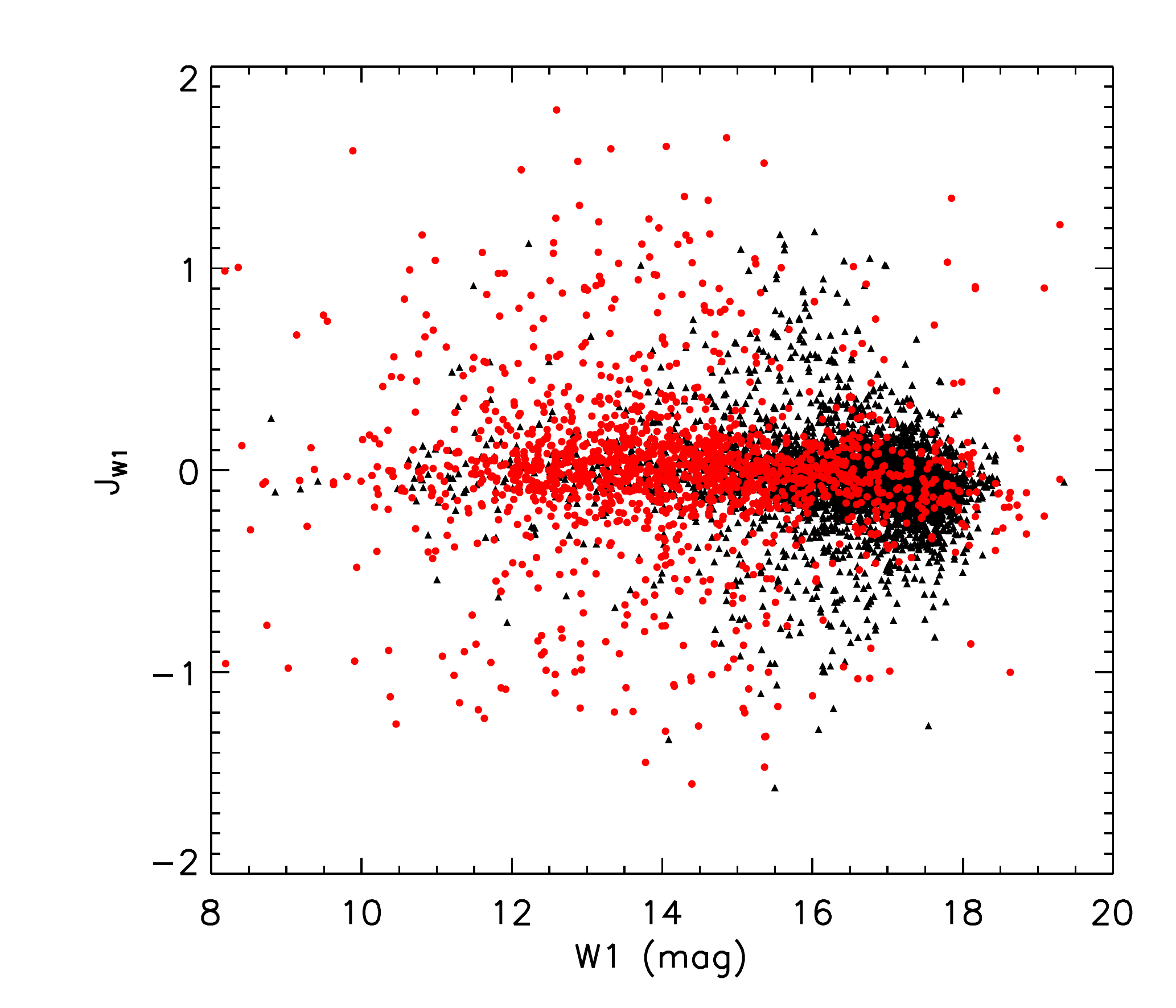}
\includegraphics[scale=0.52,angle=0]{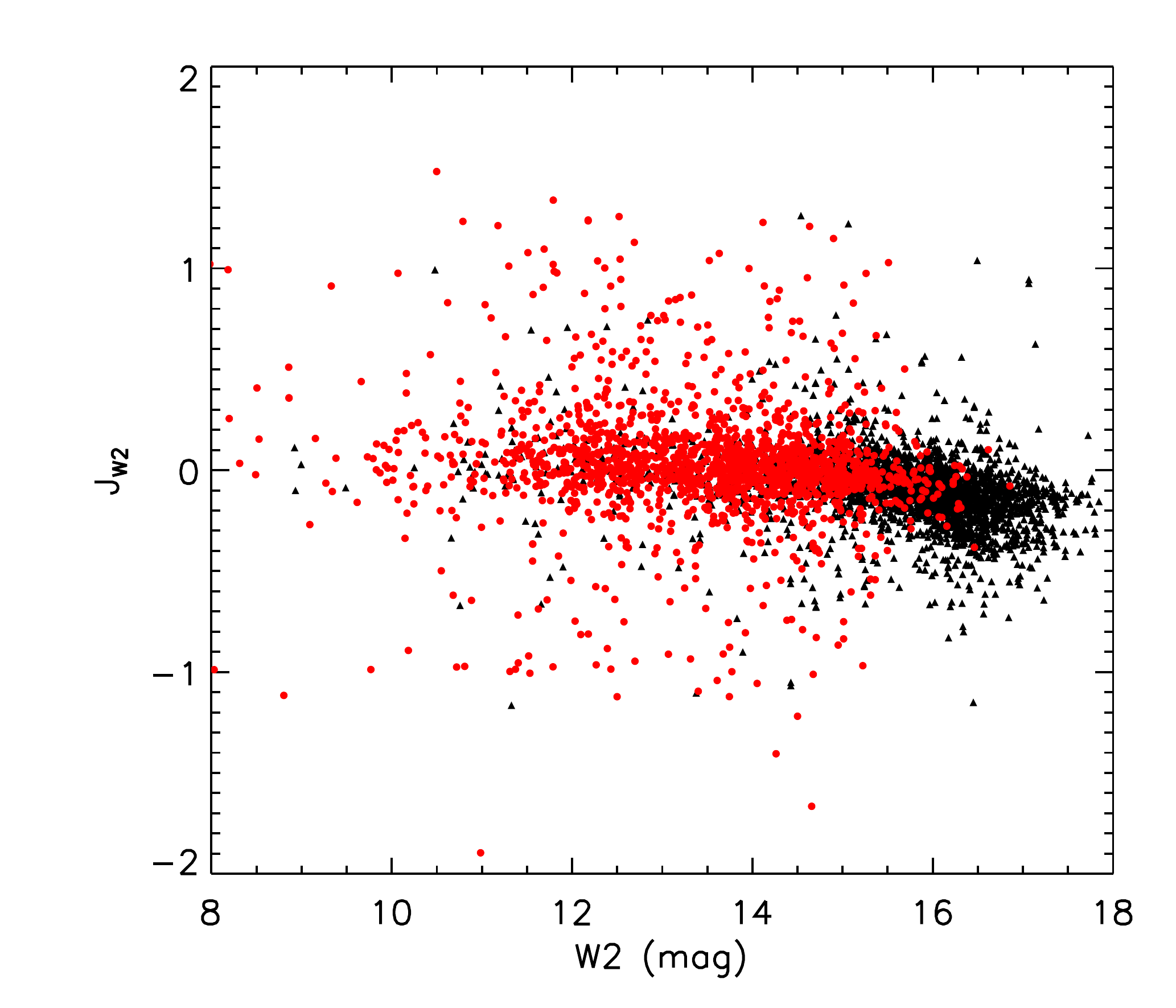}
\caption{Stetson J$_{W1}$ and J$_{W2}$ indices as a function of magnitude. Brown dwarfs are shown in red and the background sources are black.
Typical uncertainties are 0.15 and 0.21 at W2$=$12 and 16 mag, respectively. 
\label{fig2}}
\normalsize
\end{figure}

In Figure~\ref{fig2}, I show the J$_{W1}$ and J$_{W2}$ indices for the brown dwarfs and background sources as a function of magnitude.
Errors in the Stetson J Index were determined for each object by computing the index 1000 times. 
In each iteration the photometric measurements for the source were modulated by a normal distribution with the mean given by the photometric measurement in the AllWISE Source Catalog 
and standard deviation given by the uncertainty in the catalog measurements.
The median uncertainty is $\sim$0.16 for both J$_{W1}$ and J$_{W2}$. 
The dispersions in Figure~\ref{fig2} shows how the index values determined for the brown dwarfs are similar to those of the background sources.

Both the positive and negative outliers in this plot are candidate variables that must be inspected to identify real variability.
Large positive Stetson J Index values are derived from sinusoidal variations that are well sampled across the entire light curve. 
An example of this type of variability are the cepheid variables for which the Stetson J Index was defined. 
With periods on the order of a few days to a couple months, two adjacent photometric measurements in the same night will have the same sign magnitude residual, and their product will be positive; this makes the Stetson J Index positive.
For short period brown dwarfs in AllWISE, two adjacent photometric measurements are separated by $\sim$1.5 hours and the light curve is only sampled a couple times for each rotation.
In this case the Stetson J Index is determined to be negative since the sign of $\delta$ is different for the two measurements and the product is negative.
The impact of the sampling rate on the sign of Stetson J Index can hide variables with periods that beat with the WISE cadence of observation, but this can also be a tool for separating short and long term variables.
However, the dispersion in Figure~\ref{fig2} is symmetric and the single-exposure measurement uncertainties are too large to exploit this feature of the Stetson J Index.

\section{Comparing the Stetson J Index to the AllWISE Catalog Variability Flag}

\begin{figure}[!ht]
\centering
\includegraphics[scale=0.52,angle=0]{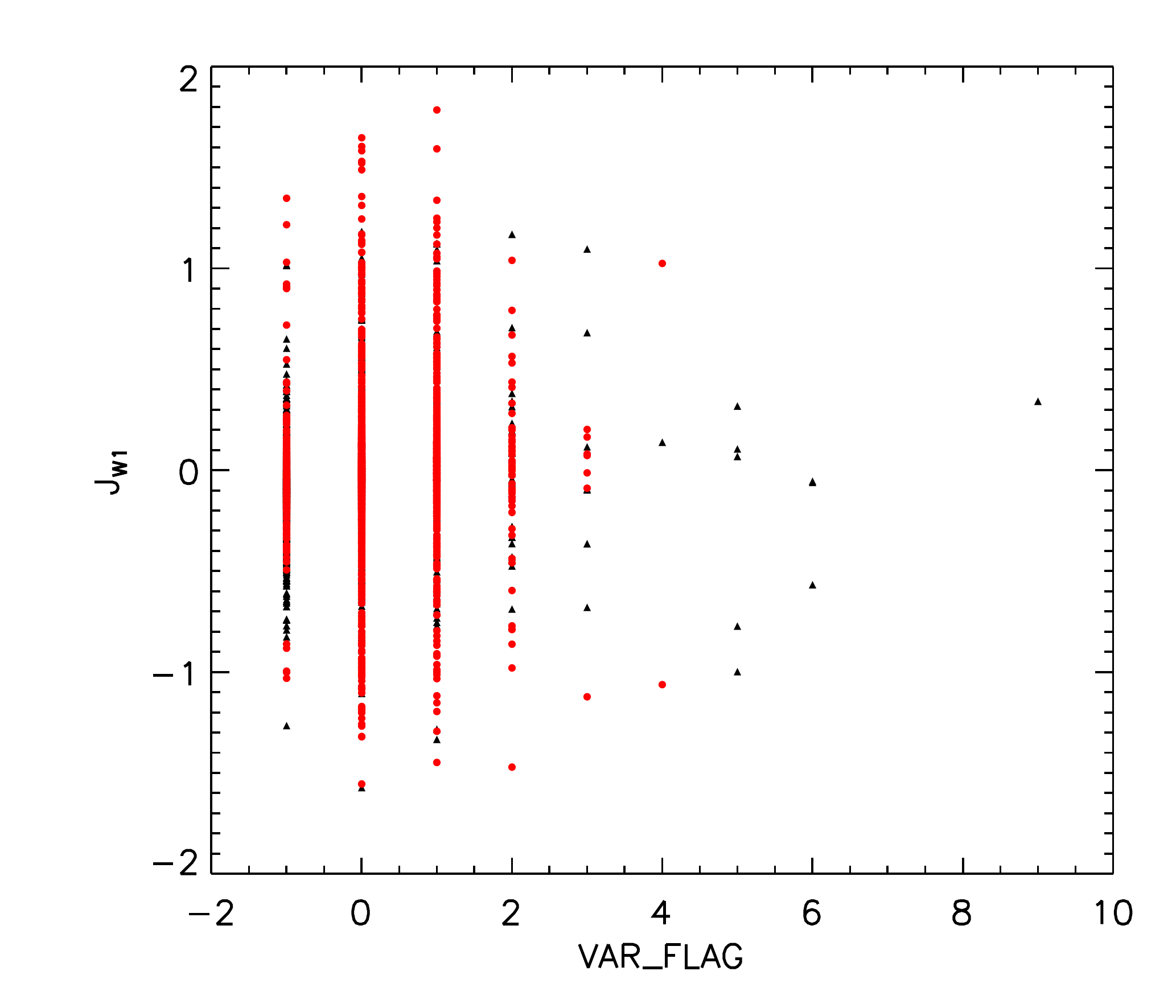}
\includegraphics[scale=0.52,angle=0]{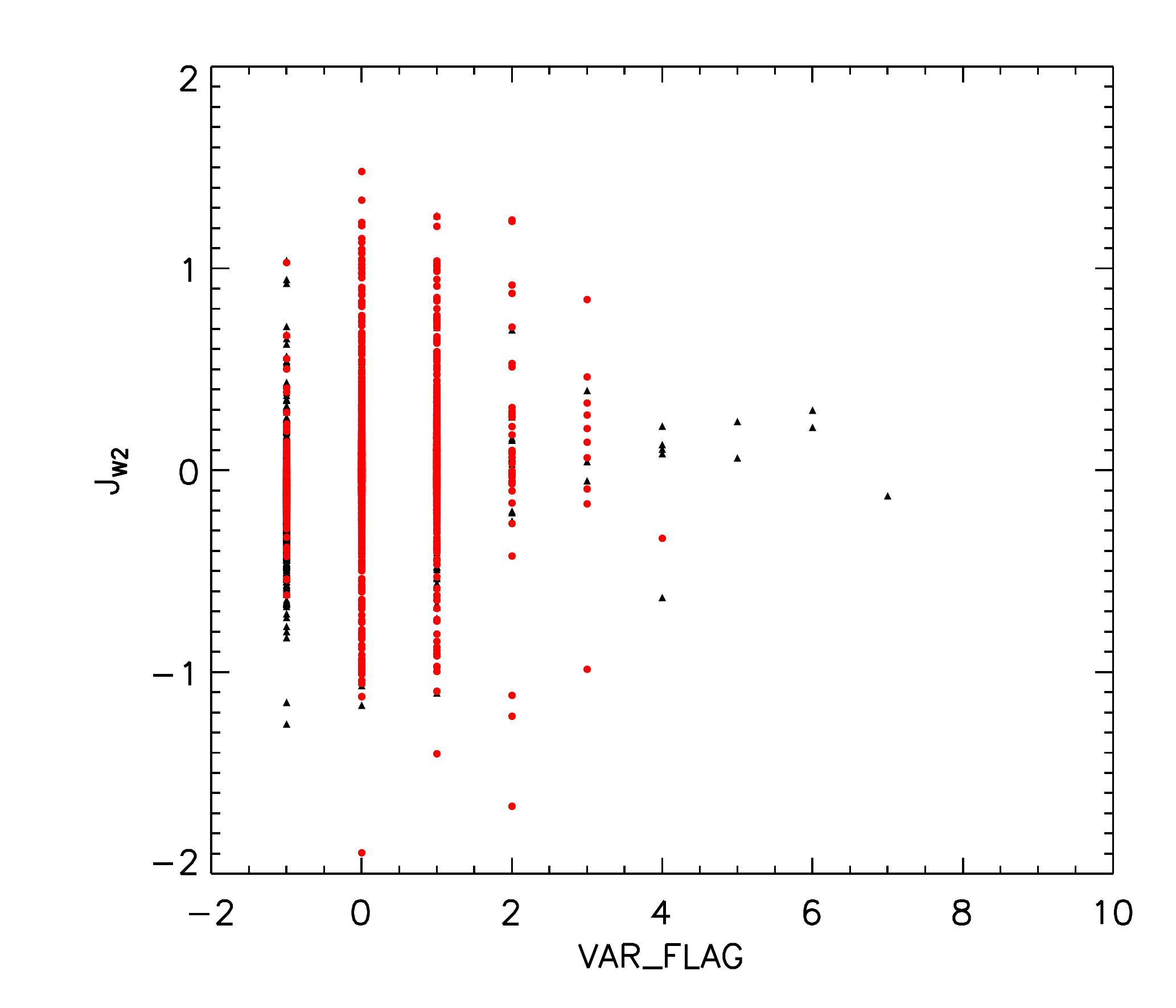}
\caption{Stetson J$_{W1}$ and J$_{W2}$ indices as a function of the AllWISE variability flag. Numbers 0 through 9 are assigned based on the dispersion in the single-exposure magnitudes relative to the background sources.
Objects with null detections have been assigned values of -1 for plotting purposes.
\label{fig3}}
\normalsize
\end{figure}

A direct comparison must be made with the AllWISE variability flag (var\_flag) to test its use for brown dwarfs.
Figure~\ref{fig3} shows the J$_{W1}$ and J$_{W2}$ indices as a function of `var\_flag'.
If the Stetson J Index and `var\_flag' were equivalent, then the largest J$_{W1}$ and J$_{W2}$ outliers would have the largest `var\_flag' values.
As described in the AllWISE explanatory supplement \citep{cutri2013}, `var\_flag' values less than ``5" are most likely not variables, and values greater than ``7" have the highest probability of being true variables.
The Stetson J Index values that I have derived are consistent with the AllWISE variability flag and most brown dwarfs are not variable within the $\sim$0.2 mag single-exposure uncertainties.

\section{Variability by Spectral Type and Subtype}

\begin{figure}[!ht]
\centering
\includegraphics[scale=0.52,angle=0]{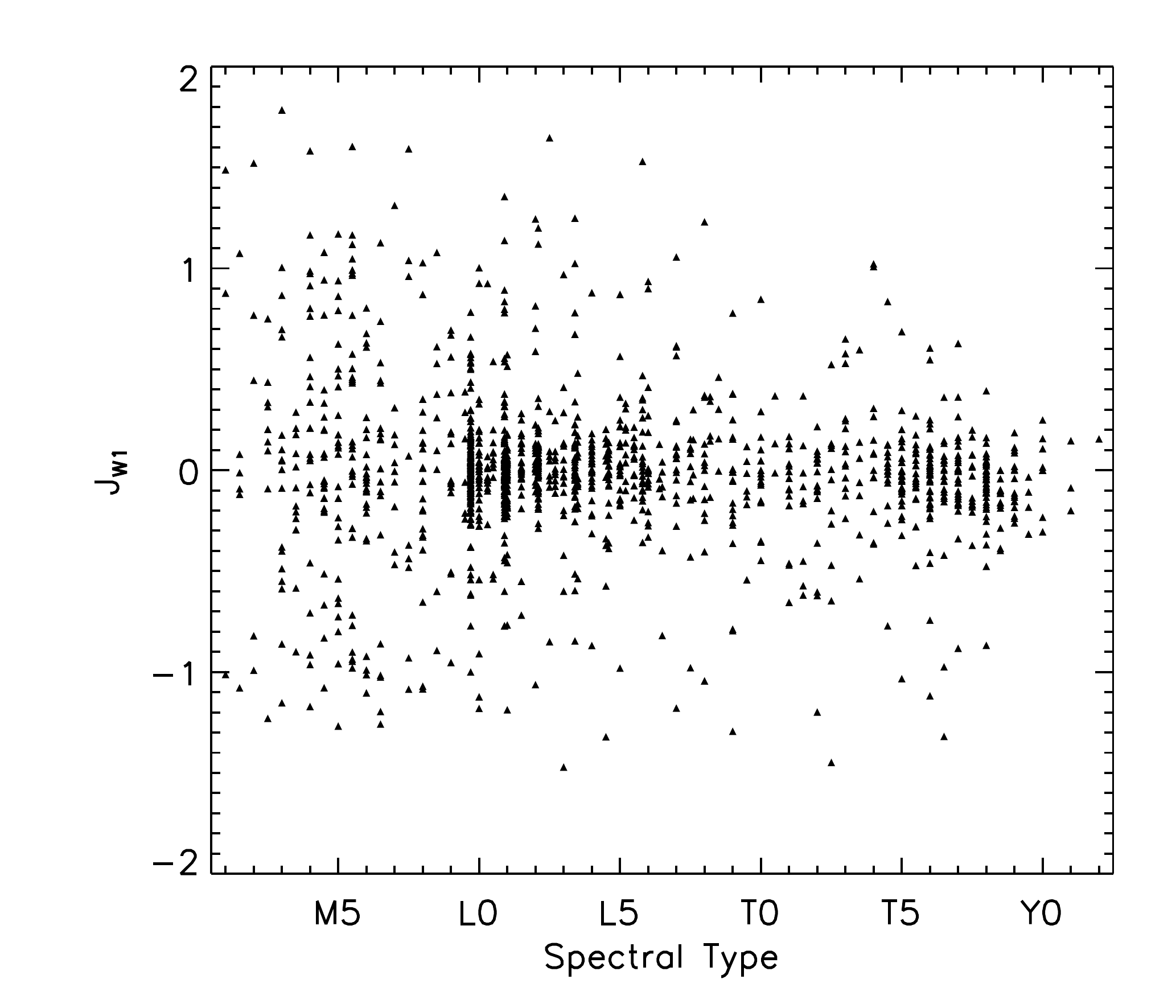}
\includegraphics[scale=0.52,angle=0]{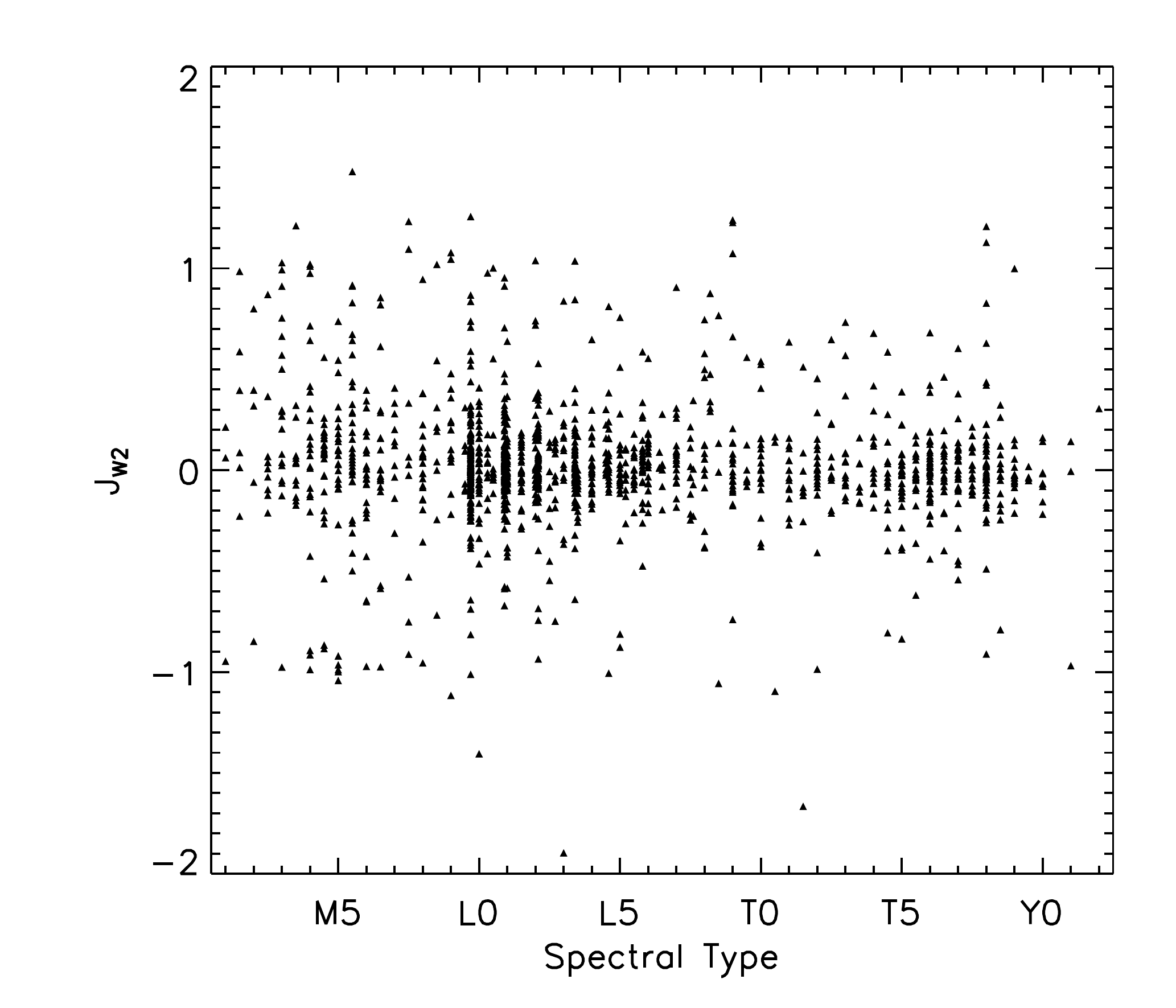}
\caption{Stetson J$_{W1}$ and J$_{W2}$ as a function of spectral type. M dwarfs have optical spectral types, L dwarfs with optical types were converted to near-infrared types, and T and Y dwarfs have near-infrared spectral types. There is no clear change in the index as a function of spectral type, and the J$_{W1}$ index dispersion is smaller for T and Y dwarfs because methane absorption in the W1 passband decreases the flux and increases the uncertainties. Typical uncertainties are $\sim$0.16. 
\label{fig4}}
\normalsize
\end{figure}

I define variability as J$_{W1}$ or J$_{W2}$ $>$ $\sigma$$_{comp}$ , where $\sigma$$_{comp}$ is the standard deviation of J$_{W1}$ and J$_{W2}$ for $\sim$5000 background stars selected to be within 30$\farcs$0 of a brown dwarf. 
For both bands $\sigma$$_{comp}$$\approx$0.3 and the average brown dwarf has a variability amplitude low enough that we can't confidently identify it in AllWISE.
Figure~\ref{fig4} shows J$_{W1}$ and J$_{W2}$ as a function of spectral type. 
M dwarfs have optical spectral types, L dwarfs with optical types were converted to near-infrared types, and T and Y dwarfs have near-infrared spectral types. There is no clear change in the index as a function of spectral type, and the J$_{W1}$ index dispersion is smaller for T and Y dwarfs because methane absorption in the W1 passband decreases the flux and increases the uncertainties.

\begin{figure}[!ht]
\centering
\includegraphics[scale=0.52,angle=0]{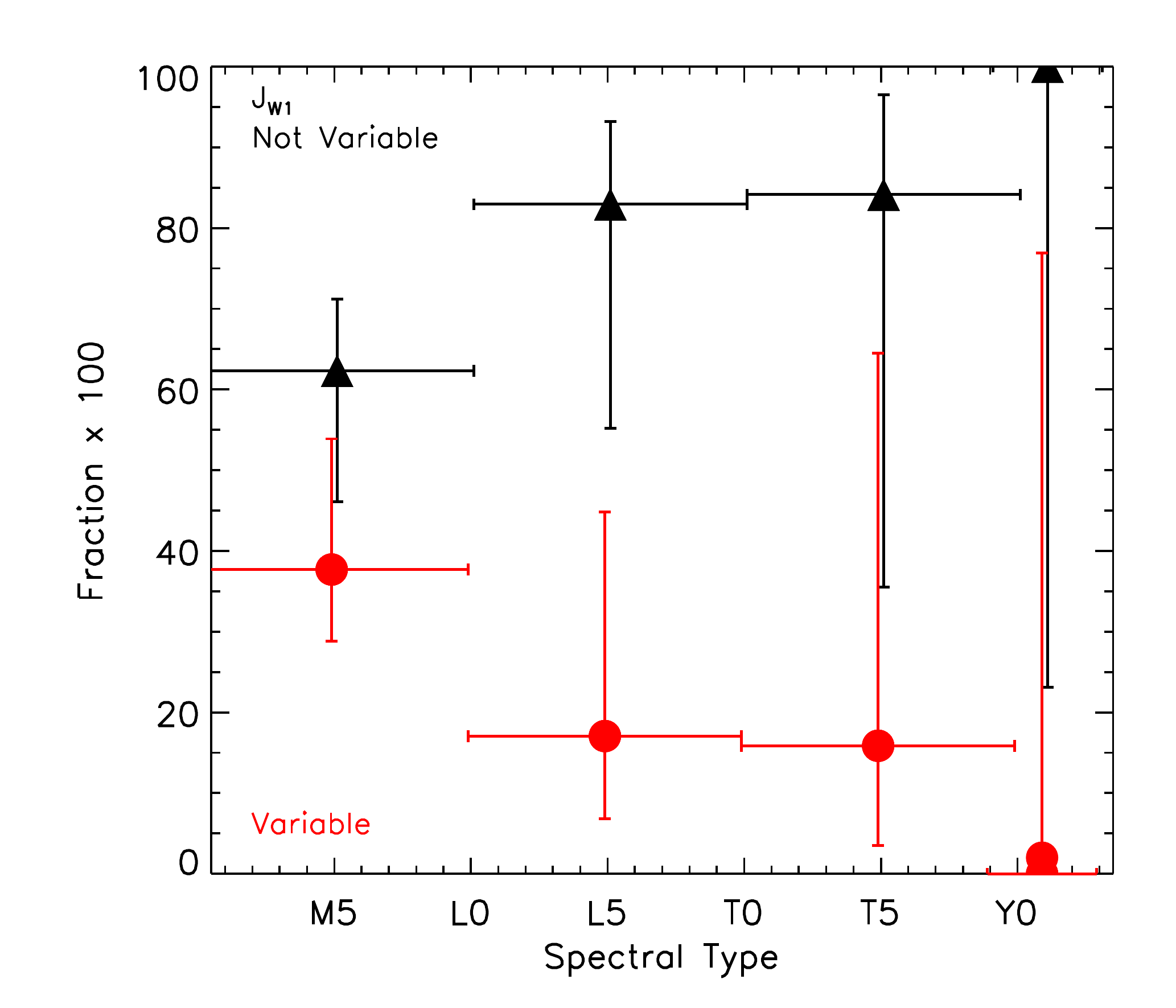}
\includegraphics[scale=0.52,angle=0]{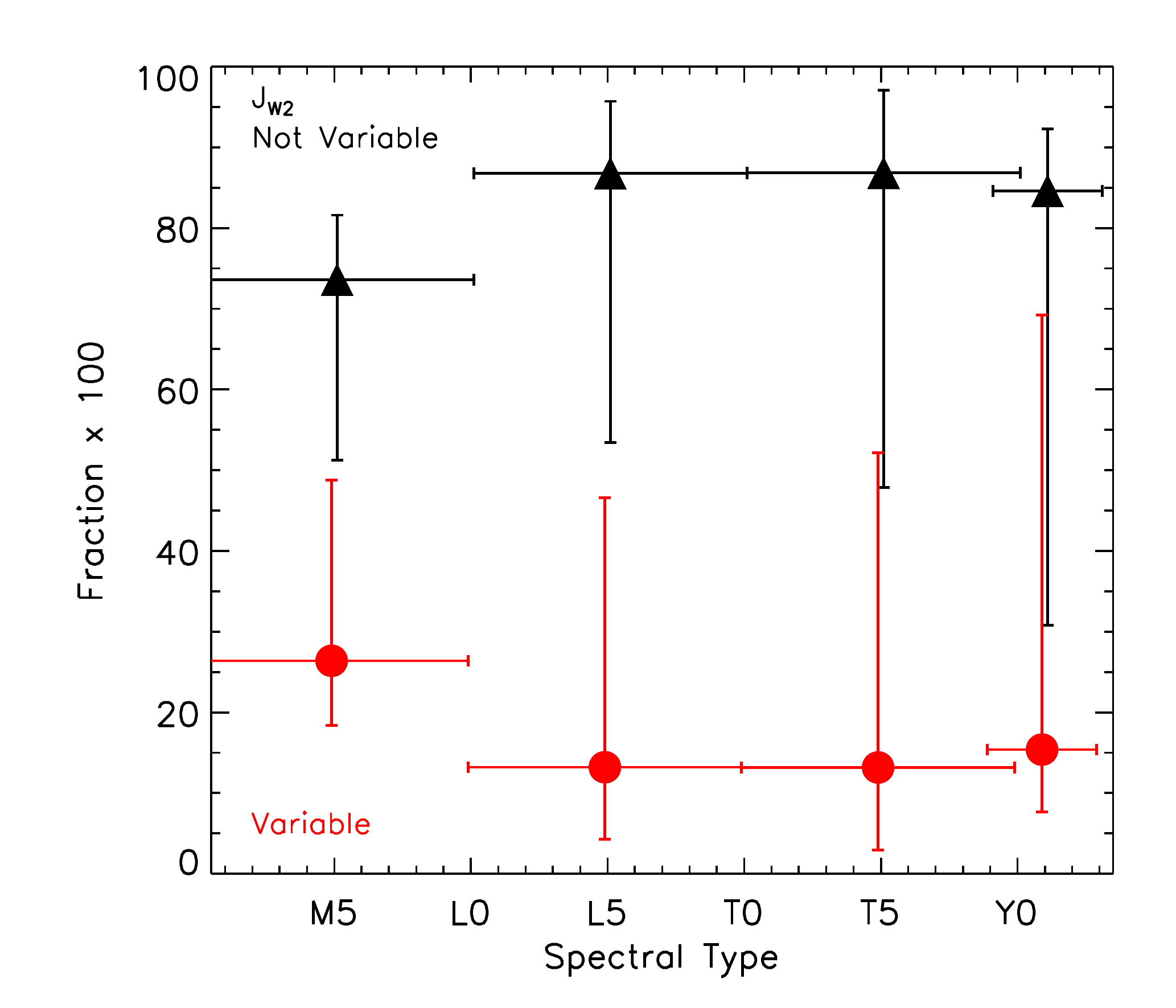}
\caption{The fraction of variable brown dwarfs binned by spectral class. Most brown dwarfs are not variable in AllWISE, but the uncertainties are large and individual sources should be checked for variability. This is especially true for large amplitude variables in the near-infrared or with known period of rotation.
\label{fig5}}
\normalsize
\end{figure}

\begin{figure}[!ht]
\centering
\includegraphics[scale=0.52,angle=0]{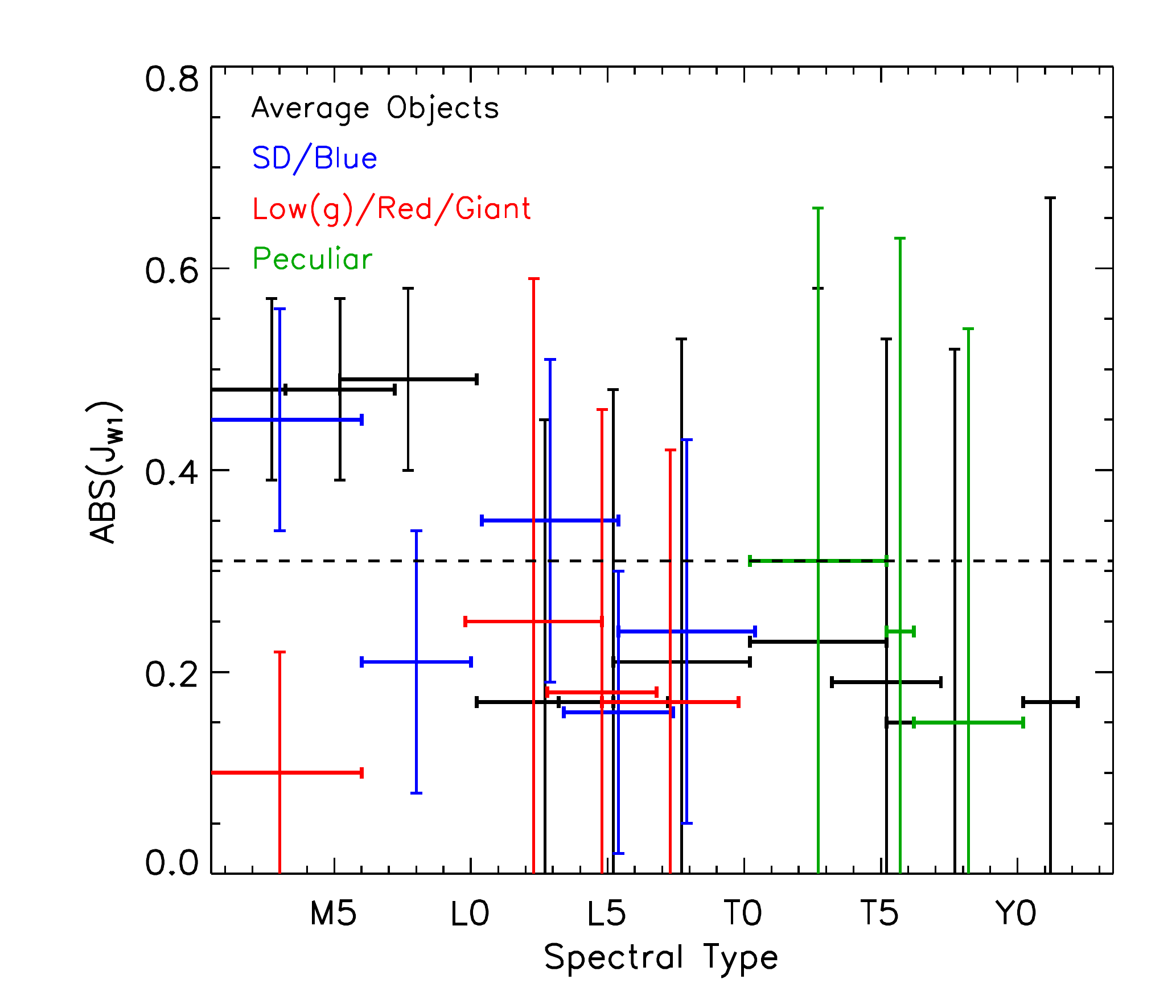}
\includegraphics[scale=0.52,angle=0]{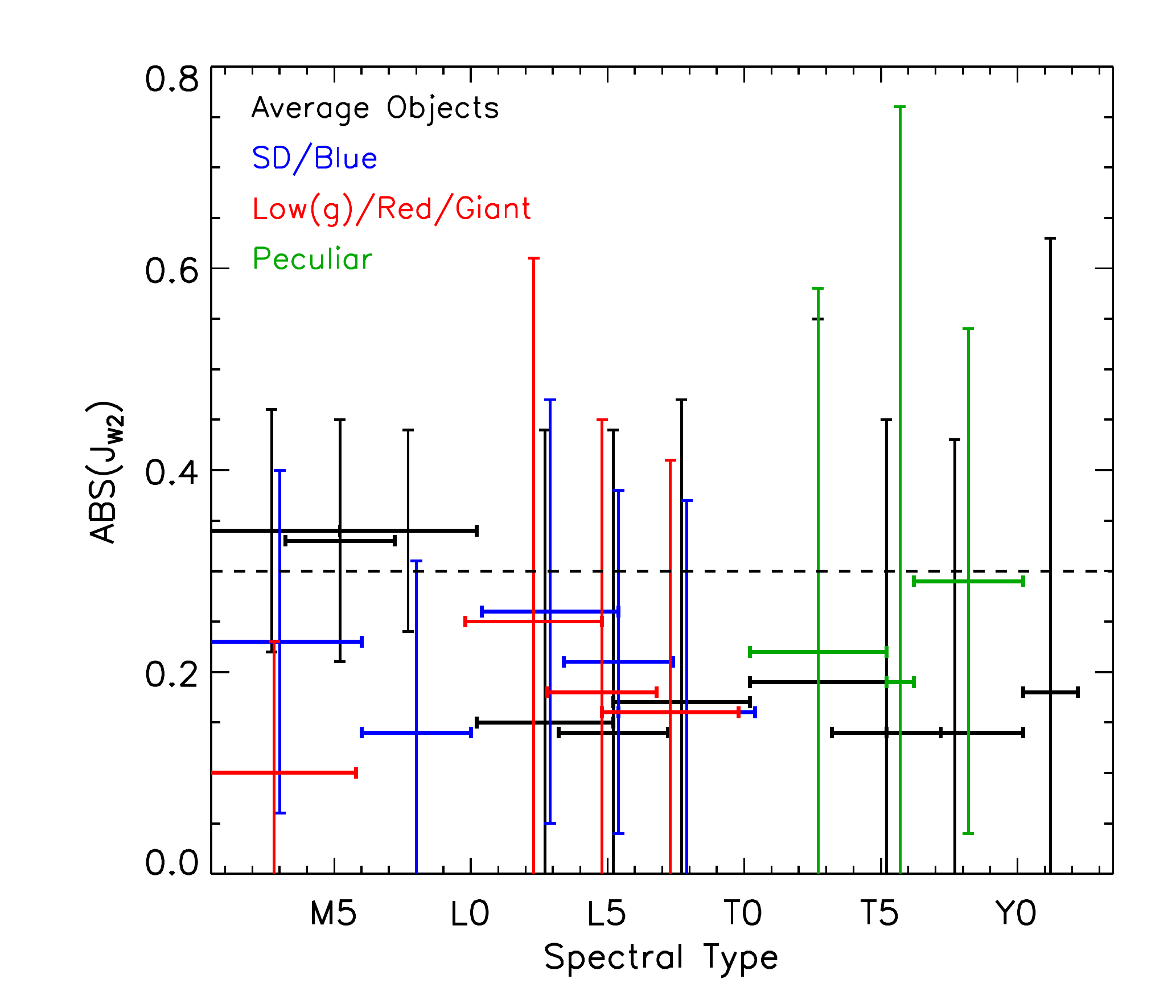}
\caption{The absolute value of J$_{W1}$ and J$_{W2}$ binned by subtype. Most of the brown dwarfs in AllWISE are not more variable than the background (dashed line). However, the large error bars on each bin show that some objects are candidate variables and should be investigated further. 
\label{fig6}}
\normalsize
\end{figure}

In Figure~\ref{fig5} I have binned the M, L, T, and Y dwarfs and computed the fraction of objects that are variable.
Sparse sampling of brown dwarf light curves by AllWISE prevents robust periodogram analysis, but the addition of periods from other observations would make this possible.
In Figure~\ref{fig6}, I plot the absolute J$_{W1}$ and J$_{W2}$ indices binned by spectral subtype. 
There are no distinguishable differences between the various subtypes, and the average M dwarf is found to be variable in AllWISE.
The uncertainties on each of the bins are large, which signals the presence of individual variable sources within the sample.

\clearpage

\clearpage

\clearpage

\clearpage

\clearpage

\section{Variability in Specific Objects}
\begin{figure}[!ht]
\centering
\includegraphics[trim=31mm 21mm 41mm 26mm, clip, scale=0.35,angle=0]{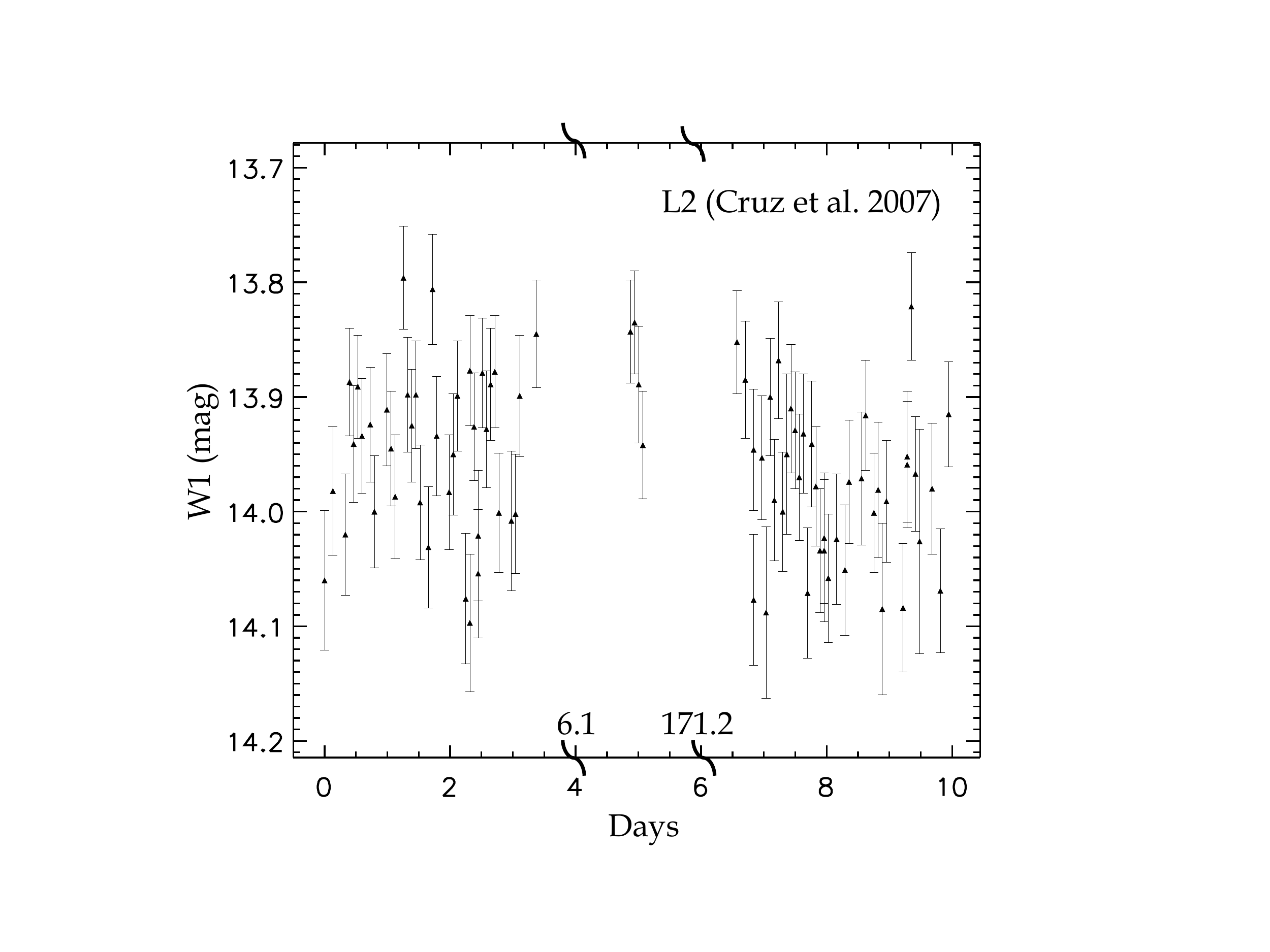}
\includegraphics[trim=31mm 21mm 41mm 26mm, clip, scale=0.35,angle=0]{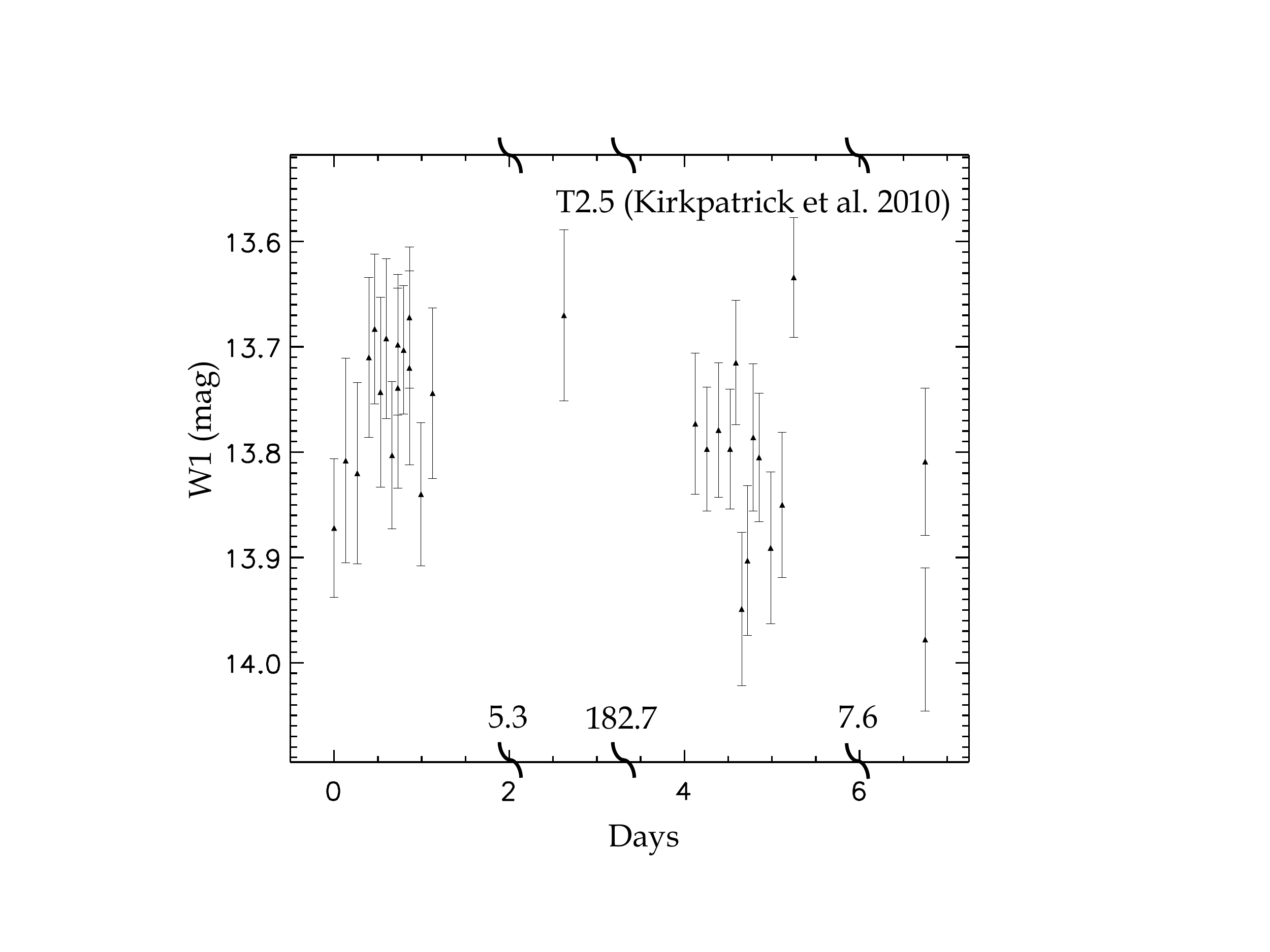}
\includegraphics[trim=31mm 21mm 41mm 26mm, clip, scale=0.35,angle=0]{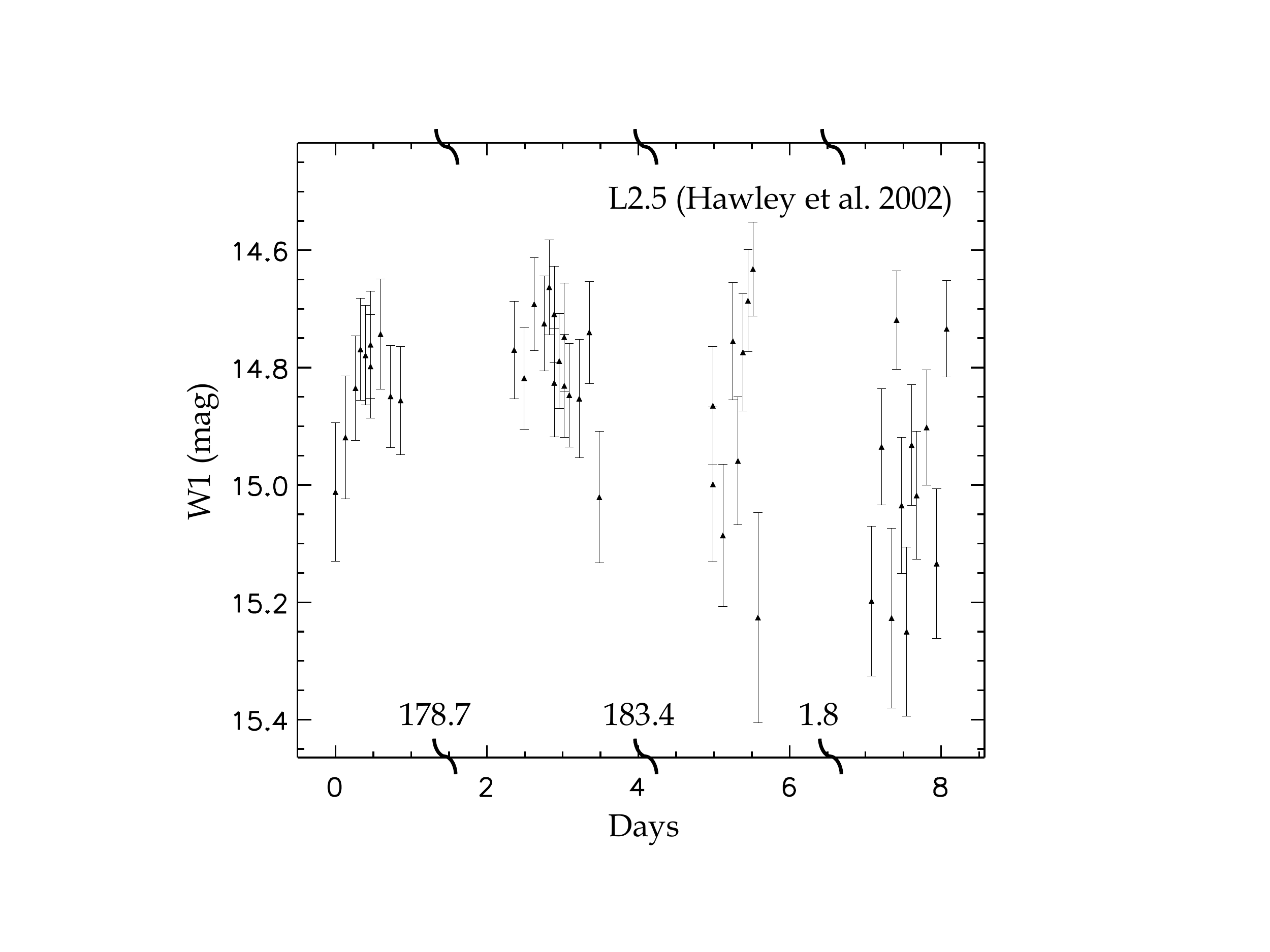}
\includegraphics[trim=31mm 21mm 41mm 26mm, clip, scale=0.35,angle=0]{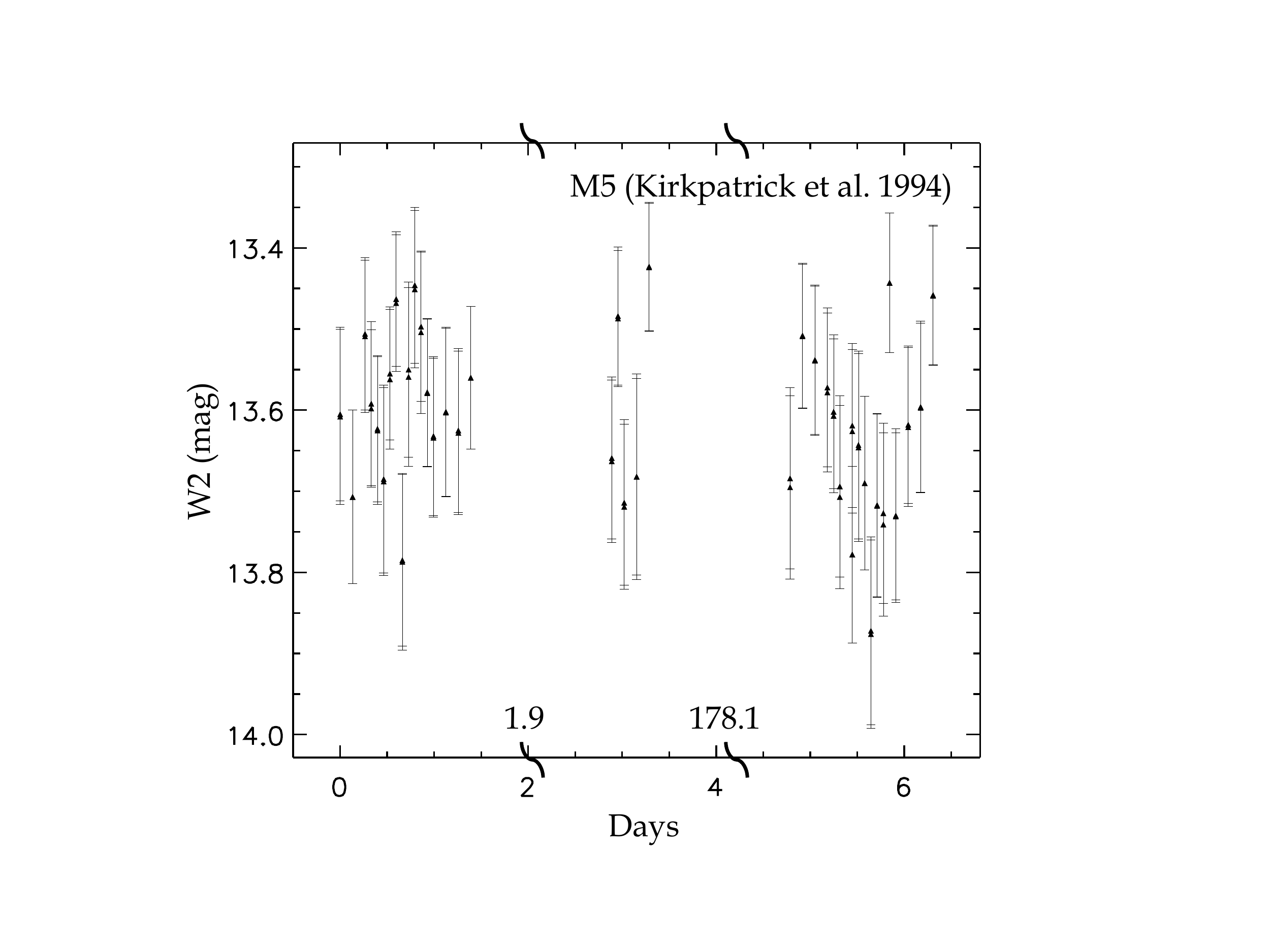}
\includegraphics[trim=31mm 21mm 41mm 26mm, clip, scale=0.35,angle=0]{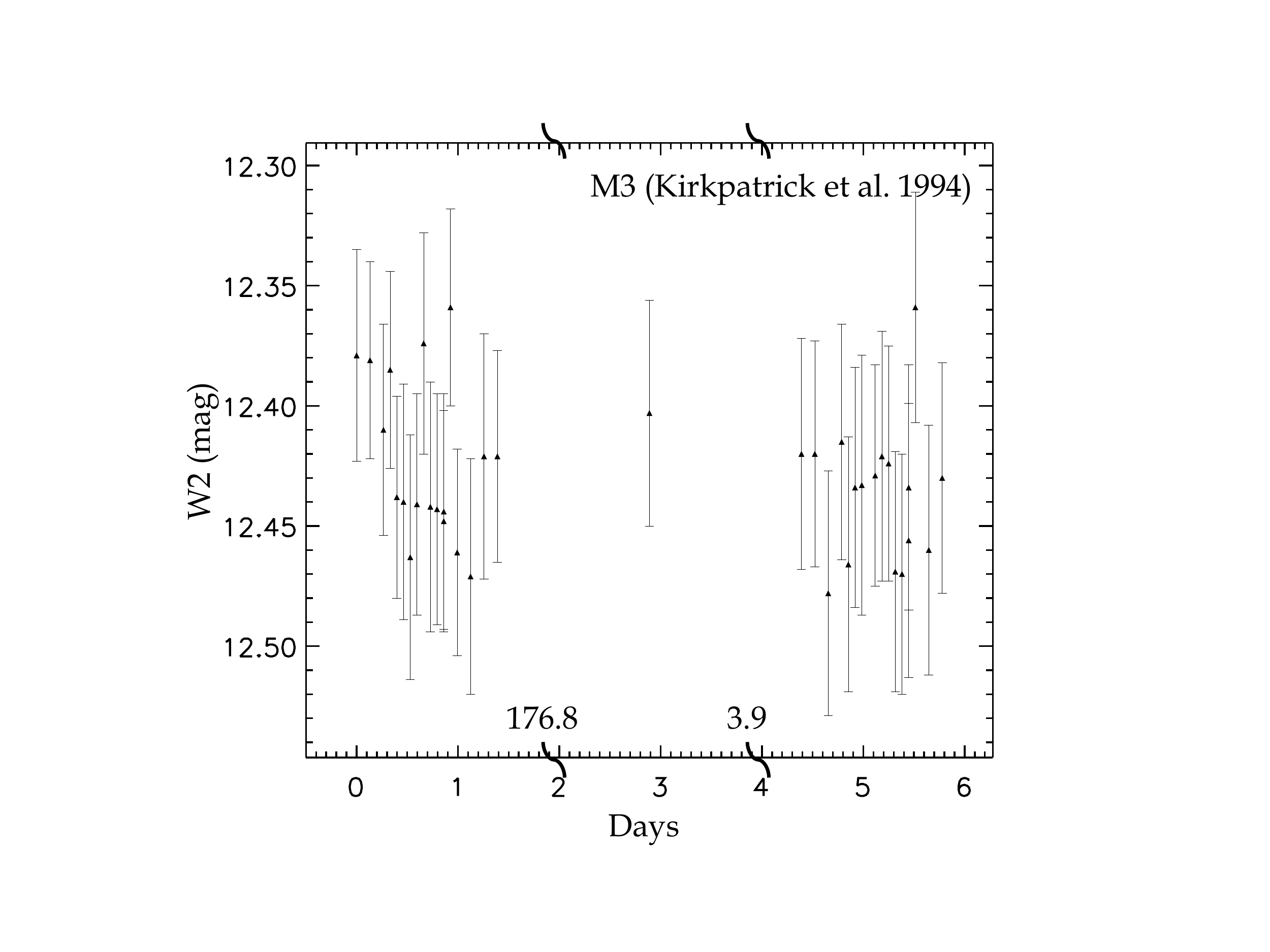}
\includegraphics[trim=31mm 21mm 41mm 26mm, clip, scale=0.35,angle=0]{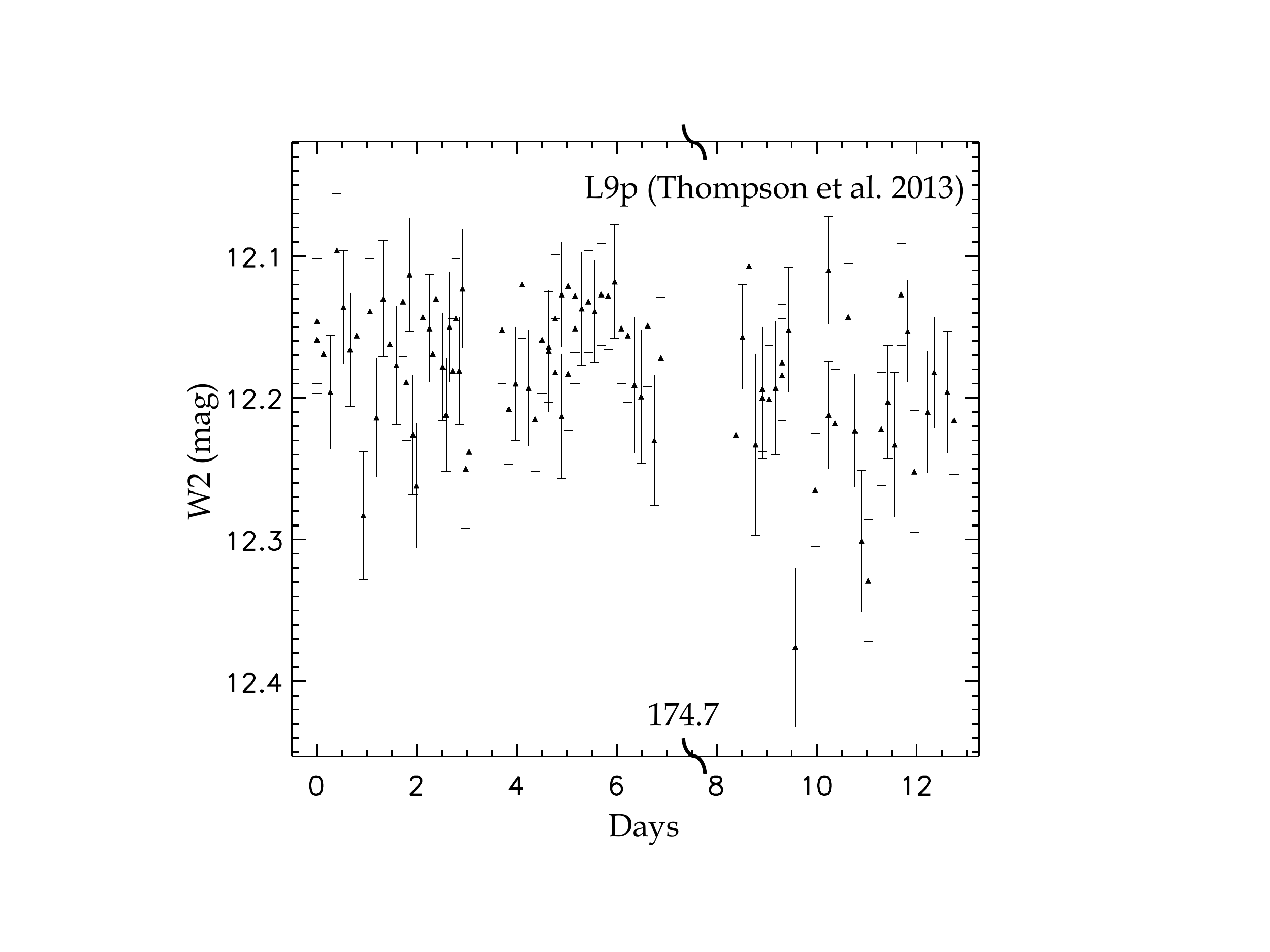}
\caption{Panel of six variable sources with spectral types and references marked. Gaps in AllWISE coverage longer than a day are compressed here for clarity.
The size of the gap, in days, is marked along the bottom axis. 
\label{fig7}}
\normalsize
\end{figure}

Single object variability is worth inspecting for your favorite source since there are true variables within AllWISE.
Figure~\ref{fig7} shows the AllWISE multi-epoch photometry for some of the objects that I have identified as variable.
There are trends in the photometry over short timescales of a few hours and over longer timescales of many months. 
The number of measurements at each epoch is determined by the proximity of the source to one of the the ecliptic poles, which have the deepest WISE coverage.

\subsection{The Nearest L, T, and Y dwarfs}
The nearest objects discovered by WISE are the Y dwarf WISE J085510.83$-$071442.5 \citep{luhman2014} and the L/T transition binary WISE J104915.57$-$531906.1 \citep{luhman2013}.
Their close proximity to Earth makes them the brightest examples of their spectral types.
Although this should produce exceptional photometry and the best test of variability, I was not able to study them in my analysis of AllWISE photometry.
WISE J085510.83$-$071442.5 was removed from the sample because it is blended with a background source in early WISE epochs, as discussed by \citet{wright2014}.
WISE J104915.57$-$531906.1 is nearly resolved by WISE, which gets it marked as an extended source in the catalog, and has a W1 magnitude $<$8 mag, which is within the saturation limits of AllWISE.
However, a number of other studies have targeted these sources \citep{gillon2013,biller2013,crossfield2014,burgasser2014}, and their variability will be an ongoing subject of study.

\subsection{The Largest Amplitude Variable T Dwarf Known}

\begin{figure}[!ht]
\centering
\includegraphics[trim=31mm 21mm 41mm 26mm, clip, scale=0.6,angle=0]{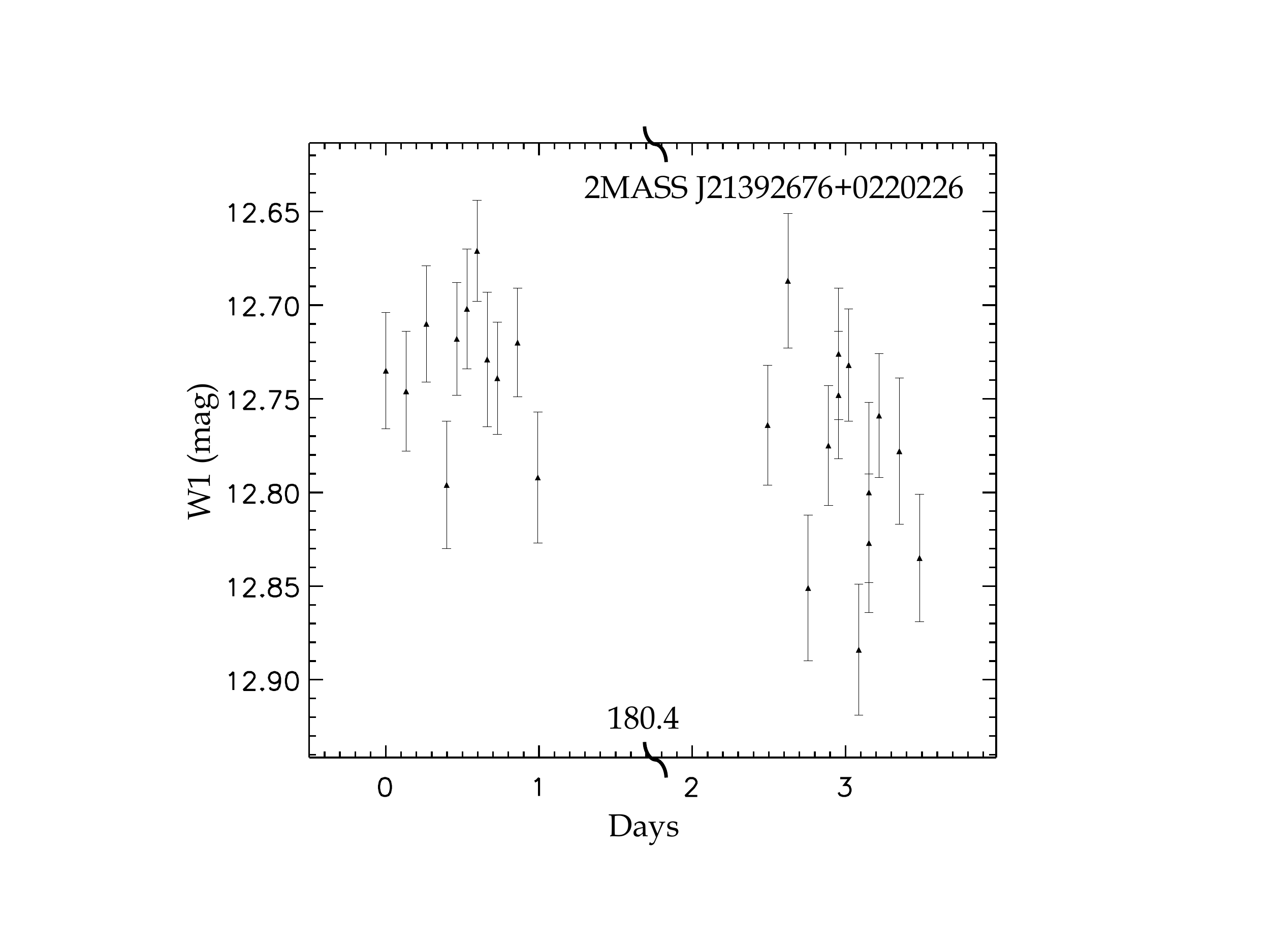}
\includegraphics[trim=21mm 21mm 41mm 26mm, clip, scale=0.6,angle=0]{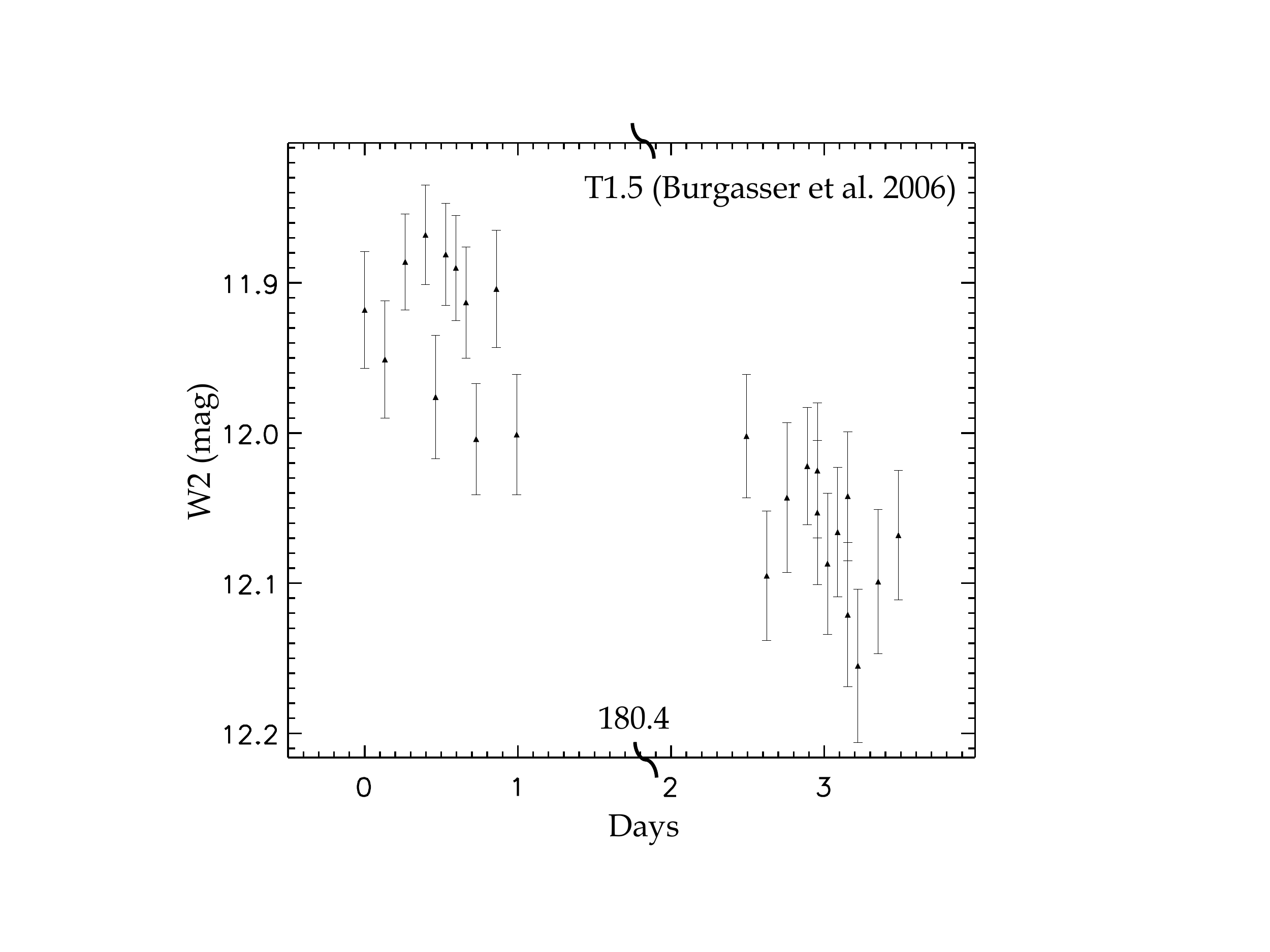}
\caption{The T1.5 dwarf 2MASS J21392676+0220226 is known to be a large amplitude variable in the J band. The AllWISE W1 and W2 multi-epoch photometry displays variability on both short (couple hour) and moderate (many month) timescales.
\label{fig8}}
\normalsize
\end{figure}

\citet{radigan2012} identify 2MASS J21392676+0220226 \citep[T1.5; ][]{burgasser2006,reid2008} as a large amplitude ($\sim$26\%) variable in the J band with a period of 7.7 hours.
They also identify long-term (10 year) changes in the J band magnitude.
In Figure~\ref{fig8} I show the W1 and W2 multi-epoch photometry for this object, which displays both short term (couple hour) and moderate-term (6 month) variability in the thermal infrared.
\citet{apai2013} present HST spectral mapping of this target and find a similar period to \citet{radigan2012} and no evidence of a phase lag at near-infrared wavelengths.
Because the W1 and W2 passbands probe some of the lowest pressure regions of T dwarf atmospheres \citep[Figure 2 of ][]{buenzli2012}, and there is no measured phase lag, temperature perturbations might dominate the variability seem in AllWISE \citep{robinson2014}.

\acknowledgments

I thank my thesis advisor, Ian McLean (UCLA), for supporting this project while I completed my dissertation.
I am also grateful to my WISE collaborators, Davy Kirkpatrick (IPAC), Michael Cushing (Univ.\ of Toledo), Chris Gelino (NExScI, IPAC), and Adam Schneider (Univ.\ of Toledo) for providing me guidance.
A number of my peers contributed to this project and I thank Sarah Logsdon (UCLA), Emily Martin (UCLA), and David Kinder (Univ. of Toledo).
Finally, I thank the entire WISE team for their hard work on exceptional data products that have only begun to be fully utilized by the community.

\clearpage
\clearpage

\end{document}